\documentclass[aip,prl,reprint,amsmath,amsfonts]{revtex4-1}

\usepackage{amssymb}
\usepackage{amsmath}
\usepackage{bm}
\usepackage{paralist}
\usepackage{graphicx}
\usepackage{enumerate}
\usepackage{tikz}
\usepackage{graphicx}
\usepackage{verbatim}
\usepackage{etoolbox}
\usepackage{nicematrix}

\usetikzlibrary{matrix,positioning,fit}

\usepackage{hyperref}
\hypersetup{
	colorlinks,
	citecolor=blue,
	filecolor=blue,
	linkcolor=blue,
	urlcolor=blue,
	pdfproducer={}
}

\let\bbordermatrix\bordermatrix
\patchcmd{\bbordermatrix}{8.75}{4.75}{}{}
\patchcmd{\bbordermatrix}{\left(}{\left[}{}{}
\patchcmd{\bbordermatrix}{\right)}{\right]}{}{}

\newlength{\Mylen}

\usepackage{ragged2e}
\usepackage{caption}
\usepackage{ragged2e}
\DeclareCaptionJustification{justified}{\justifying}
\usepackage{soul}	

\usepackage [autostyle, english = american]{csquotes}
\MakeOuterQuote{"}

\usepackage[normalem]{ulem}

\usepackage[percent]{overpic}

\usepackage{xcolor}

\usepackage{cprotect}

\begin{document} 

\title{Maximally mixing active nematics}

\author{Kevin A.~Mitchell}
\email{kmitchell@ucmerced.edu}
 \affiliation{Physics Department, University of California, Merced, CA
   95344, USA}

\author{Md Mainul Hasan Sabbir}
 \affiliation{Physics Department, University of California, Merced, CA
   95344, USA}

\author{Kevin Geumhan}
 \affiliation{Physics Department, University of California, Merced, CA
   95344, USA}
 
\author{Spencer A. Smith}
\affiliation{Physics Department, Mount Holyoke College, South Hadley, MA 01075, USA}

\author{Brandon Klein}
\affiliation{Department of Physics and Astronomy, Johns Hopkins University, Baltimore, MD 21218, USA}

\author{Daniel A. Beller}
\affiliation{Department of Physics and Astronomy, Johns Hopkins University, Baltimore, MD 21218, USA}

\date{\today}
	
\begin{abstract}
  Active nematics are an important new paradigm in soft condensed
  matter systems.  They consist of rod-like components with an
  internal driving force pushing them out of equilibrium.  The
  resulting fluid motion exhibits chaotic advection, in which a small
  patch of fluid is stretched exponentially in length.  Using
  simulation, this Letter shows that this system can exhibit stable
  periodic motion when sufficiently confined to a square with periodic
  boundary conditions.  Moreover, employing tools from braid theory,
  we show that this motion is maximally mixing, in that it optimizes
  the (dimensionless) ``topological entropy''---the exponential
  stretching rate of a material line advected by the fluid.  That is,
  this periodic motion of the defects, counterintuitively, produces
  more chaotic mixing than chaotic motion of the defects.  We also
  explore the stability of the periodic state.  Importantly, we show
  how to stabilize this orbit into a larger periodic tiling, a
  critical necessity for it to be seen in future experiments.
\end{abstract}

\maketitle

Active matter extends the scope of soft condensed matter physics to
systems far from equilibrium~\cite{Marchetti13}, with examples ranging from
bird flocks to the cellular cytoskeleton.  These materials exhibit
self-organized collective motion arising from the interplay of local
order with internal driving forces.  Though this collective motion is
often chaotic, strong confinement by walls, wells, or spherical
vesicles can render it more
predictable~\cite{Keber14,Duclos17,Opathalage19,Hardouin19,Hardouin20,Hardouin22}.
But the confinement and
control of active materials is still poorly understood.  This Letter
reports the discovery of regular time-periodic dynamics within a
well-established continuum model of active materials with nematic
order, confined to a spatially-periodic domain.  We computationally
demonstrate a method to stabilize this orbit for laboratory
applications.  Strikingly, this periodic motion maximizes the chaotic
mixing of the fluid.

\begin{figure} 
\includegraphics[width = 1\columnwidth]{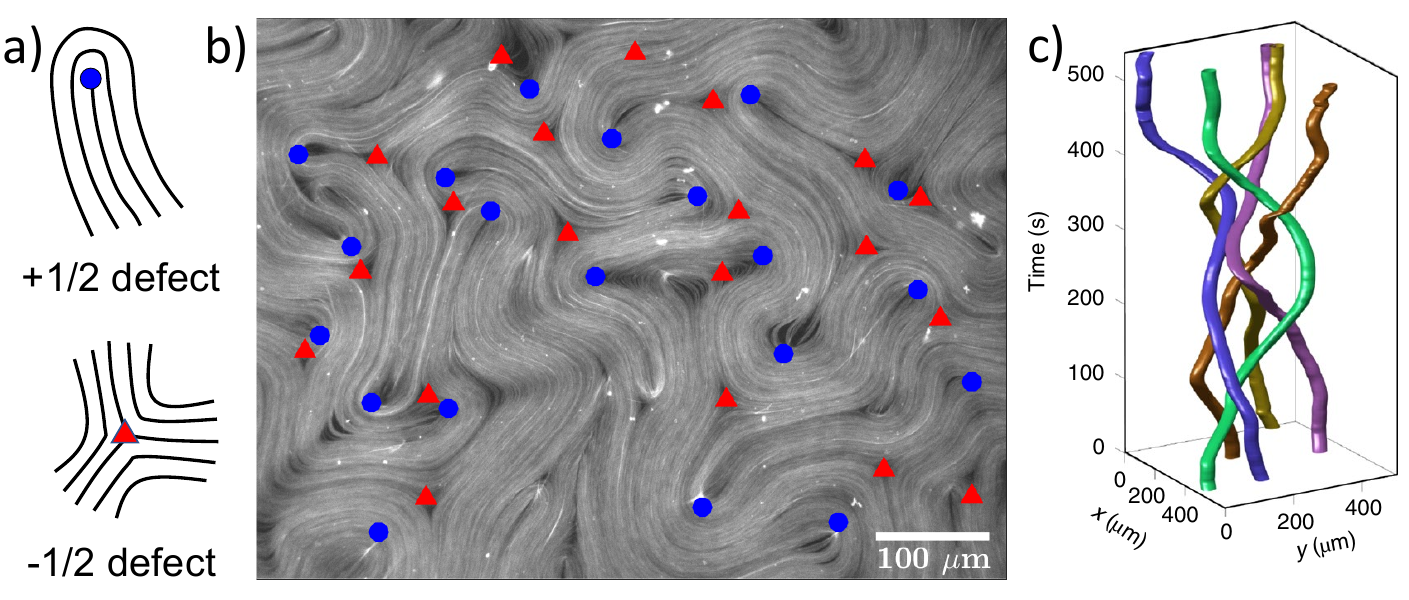}
\caption{\label{fig:MT}  a) Cartoon of defects showing nematic
  contours.  b) Fluorescence image of MT-based active nematic with
  +1/2 (circles) and -1/2 (triangles) defects. c) Space-time braiding
  of experimental +1/2 defect positions.  (Adapted from
  Ref.~\onlinecite{Tan19}.) }
\end{figure}

Active nematics are a particularly important example of active
materials.  They consist of small rodlike subunits that locally align,
forming a nematic phase, and they have a local energy source which
drives their motion.  An important model system consists of a dense
two-dimensional (2D) layer of microtubule (MT) bundles crosslinked by
kinesin molecular
motors~\cite{Sanchez12,Henkin14,Keber14,Giomi15,DeCamp15,Guillamat16,Doostmohammadi17,Guillamat17,Shendruk17,Doostmohammadi18,Lemma19,Tan19,Zhang21,Doostmohammadi22,Alert22}.
These motors hydrolyze ATP to walk along the microtubules and stretch
the bundles, injecting extensile deformations into the fluid and
driving large-scale coherent motion.  This motion is characterized by
the creation and annihilation of topological defects in the nematic
order (Fig.~\ref{fig:MT}) and the chaotic motion of these defects.
Defects have topological charges $\pm 1/2$, which are created and
annihilated in pairs to conserve topological charge.

A key goal in active nematics research is to coax them to
perform some useful control objective.  A natural
objective is to optimize their self-mixing, which could have practical
benefits in microfluidic systems, where mixing, e.g. of reagents, is
particularly difficult due to lack of turbulence~\cite{ChaoticMixingSpecialIssue}.
We characterize mixing by the amount of stretching in the
fluid, quantified by the \emph{topological entropy} $h$; in a 2D fluid
$h$ equals the asymptotic exponential stretching rate of a passively
advected material curve.  We seek to maximize this stretching over the
natural ``active'' time scale of the system, defined below.

\begin{figure} 
\includegraphics[width = 1\columnwidth]{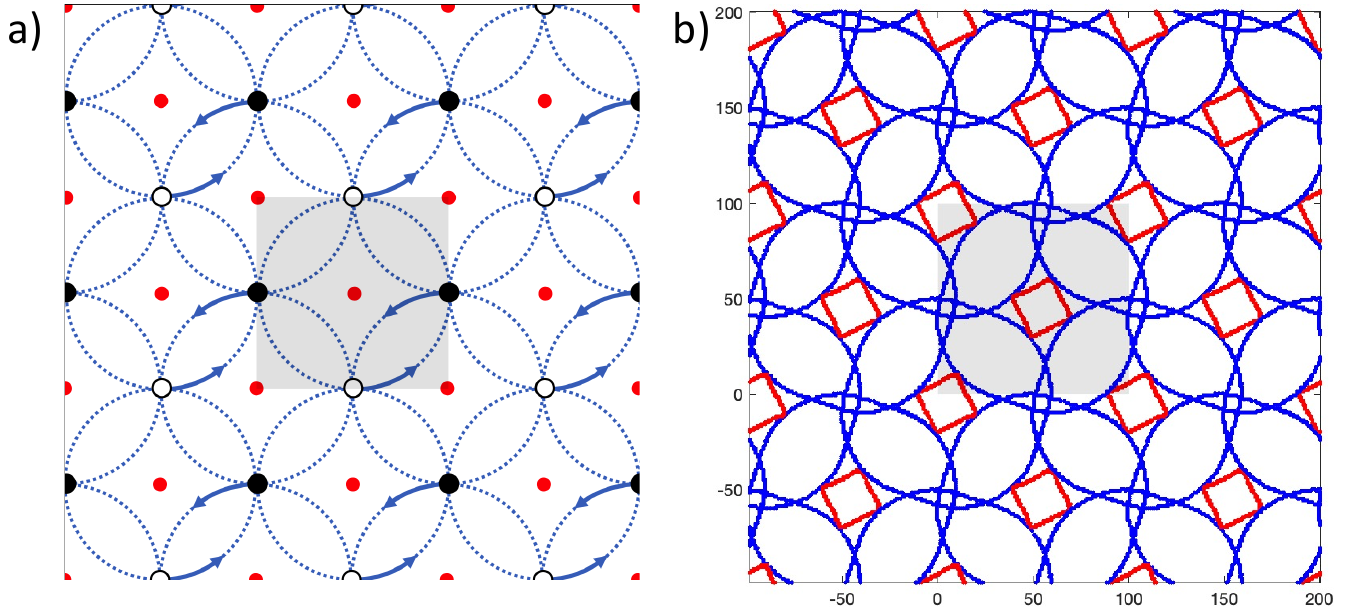}
\caption{\label{fig:braids} a) Cartoon showing the circular orbits
  (blue) of the two stirring rods (open and closed black dots) for the
  maximal mixing braid on a torus, using a periodic tiling of the
  fundamental cell (grey).  b)
  Trajectories of positive (blue) and negative (red) defects for
  simulations with $\ell_a = 3$.  The grey rectangle is the original
  integration domain, which tiles the remainder of the plane.}
\end{figure}

Reference~\onlinecite{Tan19} first applied topological entropy to active
nematics---for the MT-based active nematics, $h$ was computed
quite accurately from the space-time ``braiding'' of +1/2 defects about one
another (Fig.~\ref{fig:MT}c).  Defects act as virtual rods stirring the fluid.  Thus,
optimizing mixing reduces to coaxing the +1/2 defects into an
efficient braid pattern. 
Reference~\onlinecite{Smith22} addressed which braid to target, conjecturing, with
strong numerical evidence, that the orbit in
Fig.~\ref{fig:braids}a maximizes the topological entropy per
operation, with $h_\text{TEPO} = \log(\phi + \sqrt{\phi}) = 1.0613$,
where $\phi = (1 + \sqrt{5})/2$ is the golden ratio.  An operation is
a set of simultaneous swaps (clockwise or counterclockwise) of
neighboring defects.  Due to the numerical evidence, we call this the
maximal mixing braid.

This Letter reports that the above maximal mixing braiding state
spontaneously occurs in simulations of active nematics, when confined
to a sufficiently small square with periodic boundary conditions,
i.e.~a topological torus.  Though numerous
theoretical~\cite{Shendruk17,Norton18,Zhang20,Norton20,Samui21,Wagner22,Smith22b}
and
experimental~\cite{Keber14,Duclos17,Opathalage19,Hardouin19,Hardouin20,Hardouin22}
works have studied 2D active nematics in strongly confined geometries,
and others have studied bulk behavior (i.e. weak confinement) in
squares with periodic boundary
conditions~\cite{Thampi13,Giomi13,Giomi15}, we know of no prior study
systematically exploring the \emph{strong confinement} limit on a
square with periodic boundary conditions.

Following Ref.~\onlinecite{Giomi15}, we model the fluid velocity
$\mathbf{u}$ and nematic tensor
$\mathsf{Q} = S(\mathbf{n} \otimes \mathbf{n} - \mathsf{I}/2)$,
where $S \ge 0$ is the nematic order parameter and $\mathbf{n}$ is the director
field; a director is a unit vector with no distinction between head
and tail, i.e.~$\mathbf{n}$ and $-\mathbf{n}$ are identified.  We
numerically solve the two nemato-hydrodynamic equations as given in
Ref.~\onlinecite{Giomi15}, the first being
\begin{equation}
  \frac{D}{Dt} \mathsf{Q} = \frac{\partial}{\partial t} \mathsf{Q} +
  \mathbf{u} \cdot \mathbf{\nabla} \mathsf{Q} = \lambda S \mathsf{E} +
  [\mathsf{Q},\mathsf{\omega}] + \gamma^{-1}
  \mathsf{H},
  \label{r1}
  \end{equation}
  where $D/Dt$ is the advective derivative, $[ , ]$ is the commutator,
  $\mathsf{E} = [ (\nabla \mathbf{u}) + (\nabla \mathbf{u})^T]/2$ is
  the strain rate tensor, and
  $\mathsf{\omega} = [ (\nabla \mathbf{u}) - (\nabla \mathbf{u})^T]/2$
  is the vorticity tensor.  The \emph{flow alignment parameter}
  $\lambda$ describes the shape of the mesoscale nematogens.  Circular
  nematogens have $\lambda = 0$.  Infinitely thin needles have
  $\lambda = 1$, which we use for most of our
  simulations.  (This value of $\lambda$ is very close to that found
  in Ref.~\onlinecite{Golden23} as well.)  The first two terms on
  the right of Eq.~(\ref{r1}) derive from the passive advection of the
  microtubules.  The deviation from this behavior is given by the
  \emph{rotational viscosity} $\gamma$ and the \emph{molecular tensor}
  \begin{equation}
    \mathsf{H} = - \frac{\delta}{\delta \mathsf{Q}} F_\text{LdG}
    =  -\mathsf{Q}(A + C \text{Tr}(\mathsf{Q}^2)) + K \nabla^2 \mathsf{Q},
    \label{r3}
  \end{equation}
    which is the variation of the Landau-de Gennes (LdG) free energy
    $F_\text{LdG}$ with $C = -A>0$.  The first term describes the
    isotropic-nematic phase transition, and the second term derives
    from the elastic energy, with elastic constant $K$.

    The second dynamic equation is Navier-Stokes,
\begin{equation}
  \rho \frac{D}{Dt} \mathbf{u} = \rho \left(\frac{\partial}{\partial t}
  \mathbf{u} + \mathbf{u} \cdot \nabla \mathbf{u} \right)
  =\rho \nu \nabla^2 \mathbf{u} +
 \mathbf{F} - \nabla p,
 \label{r2}
\end{equation}
assuming incompressibility, $\nabla \cdot \mathbf{u} = 0$, with
constant density $\rho$.  The first term on the right is the viscous
drag, with viscosity $\nu$, and the third term is the force density
due to the pressure $p$.  The force density $F_i = \nabla_j \Pi_{ij}$
comes from the elastic and active stresses $\Pi = \Pi^E + \Pi^A$,
with $\Pi^E = -\lambda \mathsf{H} + [\mathsf{Q},\mathsf{H}]$ and
$\Pi^A = -\alpha \mathsf{Q}$, and where the activity $\alpha$ in the
MT system depends on ATP concentration, motor density, and
other material properties.

The bulk, i.e.~unconfined, material has two length scales, the active
length $\ell_a = \sqrt{K/\alpha}$ and nematic coherence length
$\ell_n = \sqrt{K/C}$, which characterize the average defect spacing
and defect core size, respectively.  Furthermore, the square domain
width $L$ determines the confinement.  The active time scale is
$t_a = K/(\alpha \nu)$ and corresponding velocity is
$v_a = \ell_a/t_a = \nu \sqrt{\alpha/K}$.  The dynamics is determined
by five dimensionless parameters: the flow alignment parameter
$\lambda$, the Reynolds number $\text{Re} = K/(\rho \nu^2)$, a
dimensionless rotational viscosity $\tilde{\gamma} = \gamma \nu/K$, a
LdG parameter $\tilde{C} = C/\alpha = (\ell_a/\ell_n)^2$, and the
confinement ratio $\ell_a/L$.  In simulations, we primarily use
$\lambda = 1$, $\text{Re} = 0.01$, $\tilde{\gamma} = 50$,
$\tilde{C} = 9.0$, and vary the confinement through $\ell_a$, keeping
$L = 100$.

\begin{figure} 
 \includegraphics[width = 1\columnwidth]{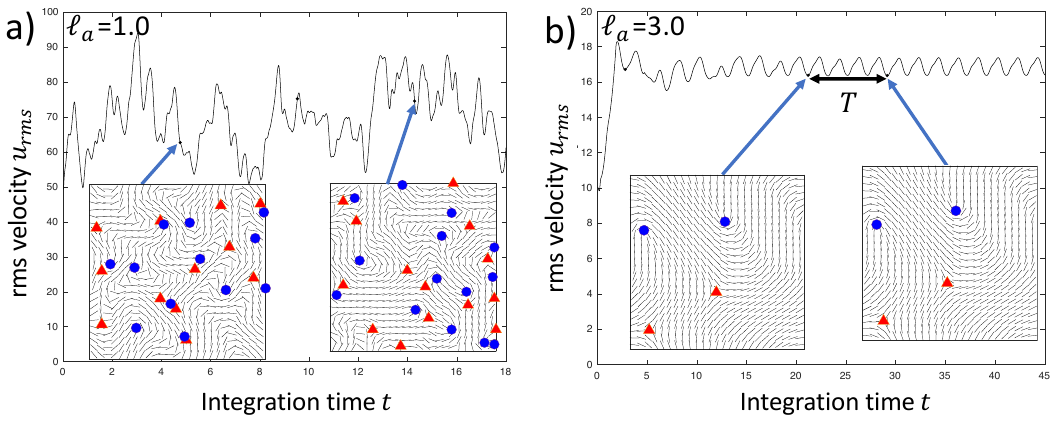}
\caption{\label{fig:sims} Simulations of active nematics for different
  values of $\ell_a$. a) $\ell_a = 1.0$.  Snapshots of the defect positions
  and director field are shown along the trace of rms velocity versus
  time.  Defect dynamics is chaotic.  b) $\ell_a = 3.0$.  Analogous plot
  for a more confined case.  Defect dynamics is periodic with period $T$.}
\end{figure}

Fig.~\ref{fig:sims}a shows snapshots of the simulation of
Eqs.~(\ref{r1}) and (\ref{r2}), with random initial condition, at
$\ell_a = 1.0$.  (See Supplemental, movie M1.)  Defects are
continuously created and destroyed, and their motion is chaotic, as
is typical for bulk behavior.  The instantaneous root-mean-square (rms)
velocity displays a chaotic time dependence.  Increasing $\ell_a$ to
$3.0$ in Fig.~\ref{fig:sims}b confines the system more tightly, with
correspondingly fewer defects.  (See Supplemental, movie M2.)  Unlike
the bulk behavior, the rms velocity quickly becomes periodic.  Two
snapshots taken at the same phase of the motion show essentially the
same director field and defect locations.  In this periodic state,
four defects trace out periodic orbits shown in
Fig.~\ref{fig:braids}b, with no creation or annihilation events.  A
key observation is that the +1/2 orbits are topologically identical to
Fig.~\ref{fig:braids}a, showing that the +1/2 defects exhibit the
maximal mixing braid.  Note that each +1/2 defect traces out a
bounded, circular shape.  The defects repeatedly encounter and revolve
around each other counterclockwise, with four such
encounters during each orbit.  The -1/2 defects trace out
a strikingly square orbit, braiding around no other defects.
Of course, by symmetry there is also a reflected orbit in which
the +1/2 defects pass each other clockwise.

The maximally mixing orbit is reminiscent of the ``Ceilidh dance''
orbit, initially observed for channel confinement by Shendruk et
al.~\cite{Shendruk17,Samui21}.  In that geometry, the +1/2 defects
aggregate along a line with half the defects moving right and the
other half left.  When opposing defects encounter one another, they
alternately pass clockwise and counterclockwise.  Several differences
distinguish the maximal mixing orbit from the Ceilidh dance: (i)
Maximal mixing defects move in a fully 2D pattern; (ii) Defect motion
is spatially bounded; (iii) Defects always pass each other in the same
sense; (iv) There are no hard-wall boundaries; (v) The orbit generates
the maximum topological entropy per operation, $h_\text{TEPO} = 1.0613$,
as conjectured in Ref.~\onlinecite{Smith22}, which is strictly
larger than that of the Ceilidh dance
($h_{\text{TEPO}} = \log(1 + \sqrt{2}) = 0.8814$)~\cite{Finn11}.
Interestingly, the Ceilidh dance is also optimally mixing, but only
under the restriction that rods are confined to a linear
arrangement~\cite{Finn11}.  In a similar manner, it has been shown
that the experimental motion of four defects on a
sphere~\cite{Keber14} is also very close to maximal mixing for that
class of defect braids~\cite{Smith22b}.  Thus, it would appear that
active nematics naturally find maximal mixing states as the system
parameters are varied.

\begin{figure}
 \includegraphics[width = 1\columnwidth]{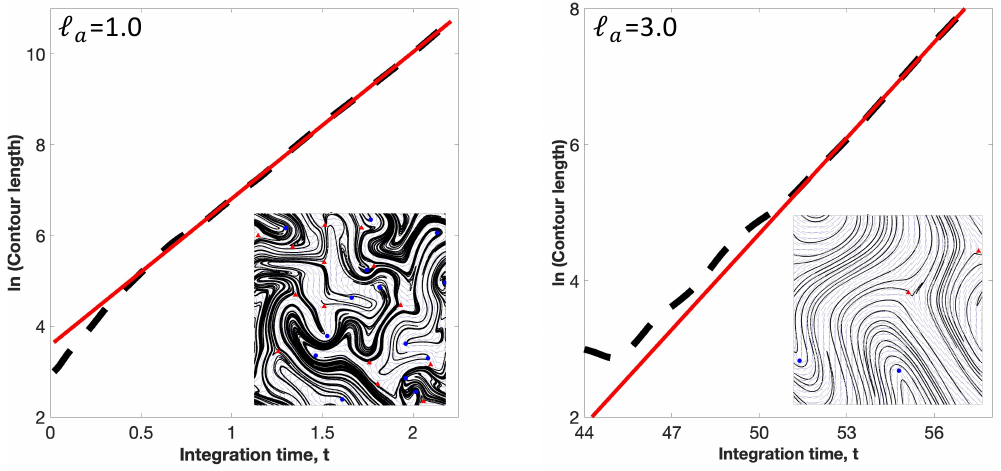}
  \caption{\label{fig:lengthvstime} Exponential growth in the contour
    length for $\ell_a = 1$ (left) and $\ell_a = 3$ (right).  Insets
    show final image of advected curves.}
\end{figure}

To compute topological entropy, we numerically advect an initial line
segment forward, recording its length versus time in the semilog plots
of Fig.~\ref{fig:lengthvstime}, whose slopes yield $h = 3.19 \pm 0.03$
($\ell_a = 1$) and $h = 0.475 \pm 0.003$ ($\ell_a = 3$).  (See also
Supplemental, movies M3 and M4.)  Here the units are reciprocal
integration time and errors are the error on the mean over four
advected curves.  Insets show final advected curves.  We define a
dimensionless topological entropy $\tilde{h} = h t_a$, which accounts
for the shift in fluid velocity with changing activity, yielding
$\tilde{h} = (1.25 \pm 0.01) \times 10^{-3}$ ($\ell_a = 1$) and
$\tilde{h} = (1.66 \pm 0.01) \times 10^{-3}$ ($\ell_a = 3$).  Thus, by
this measure, the periodic motion has more effective mixing by about
33\%.  Next, we compute the topological entropy from the E-tec
(Ensemble-based topological entropy calculation)
algorithm~\cite{Roberts19} applied to an ensemble of 1000 randomly initialized
advected trajectories.  These results agree with the line-stretching
computation to within two significant figures:
$\tilde{h} = (1.184 \pm 0.001) \times 10^{-3}$ ($\ell_a = 1$) and
$\tilde{h} = (1.665 \pm 0.003) \times 10^{-3}$ ($\ell_a = 3$).  Since
E-tec is more efficient than the line-stretching algorithm (which
becomes exponentially more expensive in time), we use E-tec for the
remainder of the paper.

\begin{figure}
\includegraphics[width = 1\columnwidth]{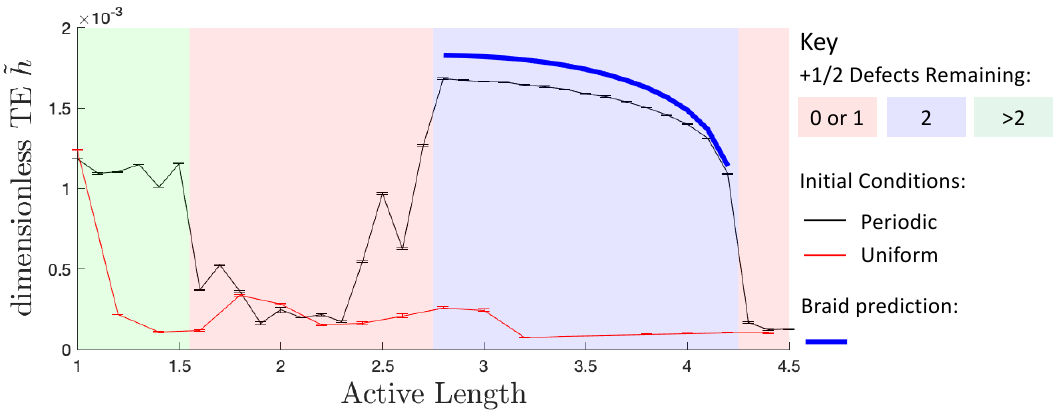}
  \caption{\label{fig:TEvsL} Topological
    entropy $\tilde{h}$ versus $\ell_a$.  The initial $\mathsf{Q}$-tensor for 
    black curve is a snapshot of the periodic state at $\ell_a = 3$,
    $\tilde{\gamma} = 50$, whereas the initial tensor for red curve is 
    nearly constant.   Red shading denotes intervals with 0 or 1
    +1/2 defects remaining.  Blue denotes two remaining and green
    three or more.}
\end{figure}

To explore the \emph{transition} to periodic behavior,
Fig.~\ref{fig:TEvsL} records $\tilde{h}$ for $1 \le \ell_a \le 4.5$.
The black curve uses an initial $\mathsf{Q}$-field taken from the
periodic state at $\ell_a = 3$.  The system either remains in a nearby
stable periodic state or departs from it.  The blue band is the
interval where the system retains two +1/2 defects for the duration of
the simulation.  Within this band, the periodic orbit is stable, or at
least sufficiently close to stable, that it remains mostly periodic
over the course of the simulation.  The topological entropy drops
precipitously on either side of the blue band, due to the periodic
state either disappearing or becoming sufficiently unstable.  On
either side of the periodic band, the system converges to a steady
state with no defects and small, nearly constant, velocity (red band).
Technically, for a small number of $\ell_a$ values in this interval, a
pair of opposite defects remained at the end of the integration, but
such a lone pair is expected to eventually annihilate when integrated
further.  At $\ell_a \lesssim 1.5$ the final state has chaotic defect
motion with three or more defects (green band) and an intermediate
value of $\tilde{h}$.

The dimensionless topological entropy of the maximal mixing braid is
$\tilde{h}_\text{max} = \log(\phi + \sqrt{\phi})/(\tilde{T}/4)$, where
$\tilde{T}$ is the period of defect motion in units of $t_a$.  This produces the
blue curve in Fig.~\ref{fig:TEvsL}, which closely tracks $\tilde{h}$,
albeit at a slightly larger value.  We attribute this difference to
the positive defects moving slightly faster than the surrounding
fluid.  For example, at $\ell_a = 3.0$ the rms difference between
defect and fluid velocities is about 34\%.  This fact is also seen by
the defect passing through the advected curves.  (See Supplemental
movies M3 and M4.)   This difference in
velocity is also seen in experiments as the MT bundles fracturing when
bent too far~\cite{Sanchez12}.

\begin{figure}
\includegraphics[width = 1\columnwidth]{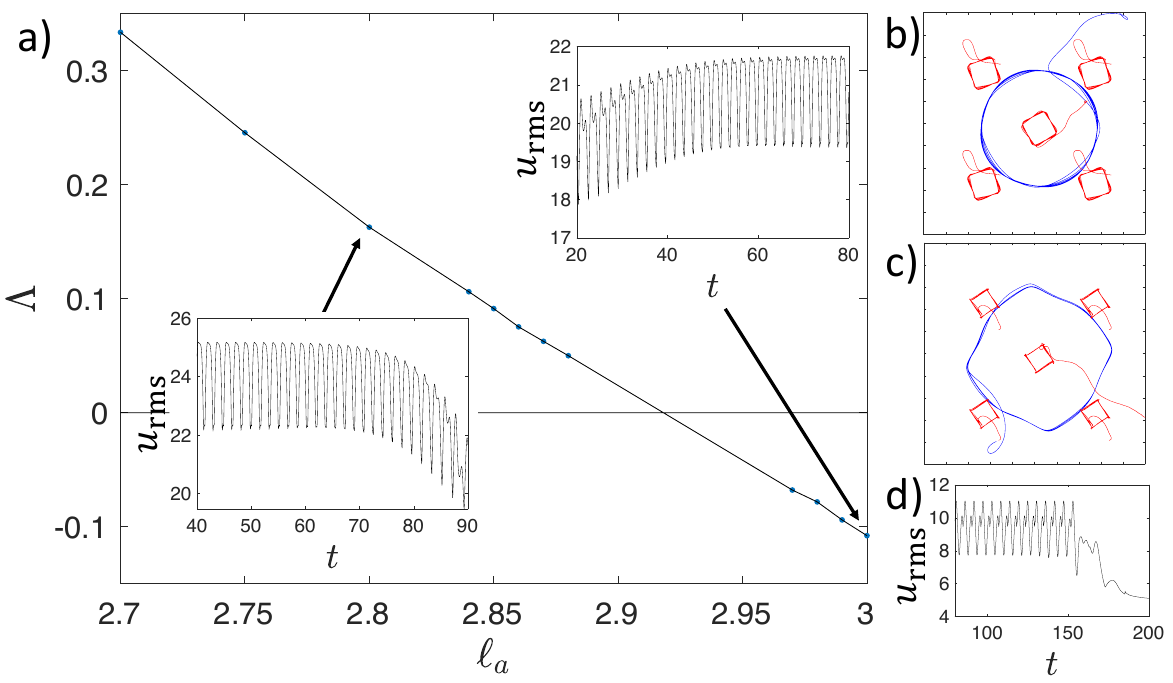}

\caption{\label{fig:LE} a) The Lyapunov exponent of the periodic
  orbit.  Insets show $u_\text{rms}$ vs. integration time for positive and
  negative $\Lambda$. b) An orbit diverging from the periodic orbit
  for $\ell_a = 2.8$.  c) Same as (b) for $\ell_a = 4.15$.  d) $u_\text{rms}$
  vs. time for orbit in (c). }
\end{figure}

To clarify the bifurcation at the left edge of the blue band in
Fig.~\ref{fig:TEvsL}, we compute the Lyapunov exponent $\Lambda$ of
the periodic solution by measuring how quickly a nearby state diverges
from it (for positive $\Lambda$) or converges to it (for negative
$\Lambda$).  See Fig.~\ref{fig:LE}a.  The Lyapunov exponent
transitions from negative to positive as $\ell_a$ decreases, passing
through zero at $\ell_a = \ell_c \approx 2.92$.  The linear trend in
$\Lambda$ and the fact that no nearby stable periodic orbit exists
below $\ell_c$, suggests that this is a subcritical pitchfork
bifurcation~\cite{Strogatz94} (and certainly not a saddle-node or
supercritical pitchfork bifurcation).  Pitchfork bifurcations
generically occur for systems with a discrete symmetry whose square is
the identity, as for certain translational, rotational, and mirror symmetries
of Fig.~\ref{fig:braids}b.  Thus, we suspect that there is a pair of
nearby unstable symmetrically related periodic solutions for
$\ell_a > \ell_c$, which collide with the stable periodic orbit at
$\ell_c$ to drive it toward instability.

For a given $\ell_a < \ell_c$, a +1/2 trajectory diverges
``adiabatically'' from the periodic state.  As each +1/2 defect swaps
past its partner, it is nudged slightly until it eventually
annihilates with a -1/2 defect.  (See Fig.~\ref{fig:LE}b and
Supplemental movie M5.)  Since each nudge is small, the periodic
oscillations in $u_\text{rms}$ persist, but with a gentle downward
trend (left inset of Fig.~\ref{fig:LE}a.)  The same trend is seen in
reverse for $\ell_a > \ell_c$, as the nudges stabilize the orbit
(right inset of Fig.~\ref{fig:LE}a.)

Unstable orbits near the right edge of the blue band behave very
differently  (Fig.~\ref{fig:LE}c.)  First, the +1/2 orbit has
developed four kinks, each coming from a head on collision between the
two +1/2 defects.  (See Supplemental movie M4.)  Eventually, small
perturbations lead to a hard scattering event that nearly
instantaneously pushes the +1/2 defect onto an entirely different
path, causing it to annihilate with a -1/2 defect.  The orbit changes
``nonadiabatically''; $u_\text{rms}$ remains periodic until suddenly
breaking down (Fig.~\ref{fig:LE}d).
  
The active nematic can exhibit bistability, with the final dynamics
dependent on the initial conditions.  The red curve in
Fig.~\ref{fig:TEvsL} is the entropy $\tilde{h}$ resulting from a
nearly uniform initial director field.  The bump seen in the black
curve due to the maximal mixing orbit is now absent.  Furthermore,
the jump in entropy due to chaotic trajectories only occurs for the
lowest value of $\ell_a$.

To understand how the shape of the nematogen affects our results, we
varied $\lambda$, keeping $\tilde{\gamma}$, $\tilde{C}$, Re fixed and
varying $\ell_a$.  We found that $\lambda$ needed to be sufficiently
large, roughly greater than or equal to 0.6 for the periodic orbit to
be visible (Supplemental Fig.~1).
We also explored how our results varied with the
rotational viscosity $\tilde{\gamma}$.  See the Supplemental Material
for details.

\begin{figure} 
\includegraphics[width = 1\columnwidth]{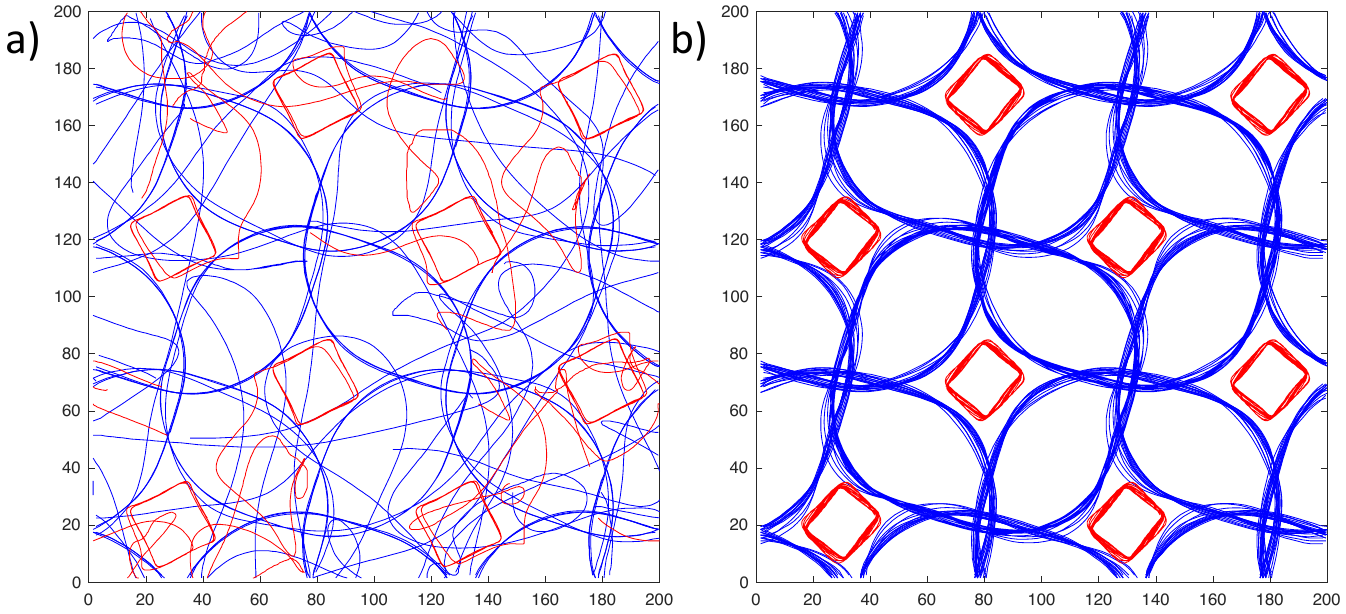}
\caption{\label{fig:posts} a) Simulation ($\ell_a = 3$) on a square
  domain having twice the width as before, with periodic boundary
  conditions.  Maximal mixing solution is no longer stable as seen by
  negative (red) and positive (blue) defect trajectories.  b)
  Identical simulation except eight control points are added following
  the negative defects.   }
\end{figure}

Recall that the maximal mixing solution relies on tight confinement
within a flat square with periodic boundary conditions; this setup is
unphysical in the lab.  So, for this solution to be experimentally
seen, it would need to be stable when periodically tiled over an
experimental domain large enough that boundary effects could be
ignored.  To test this, we ran simulations on a square with twice the
width, using an initial Q-tensor consisting of a 2x2 tiling of the
original periodic state plus a small perturbation that breaks the
discrete translational symmetry.  The computation showed
that the periodic solution was unstable (See Fig.~\ref{fig:posts}a and
the left side of Supplemental movie M7.)  Thus, this periodic motion
must be stabilized to be seen in the lab.  We focus on a simple
method that introduces local control points in the fluid; at a control
point, the potential in the LdG free energy inverts at the
origin to become a well with a single minimum.
This induces a local phase transition to the isotropic state at each
control point.  (For details on the modification to the LdG free
energy, see the Supplement.)  Our computations show that these
control points indeed successfully stabilize the maximal mixing state
(Figure~\ref{fig:posts}b and the right side of Supplemental movie M7.)  The simulation
in Fig.~\ref{fig:posts}a (without control points) uses exactly the
same initial conditions as Fig.~\ref{fig:posts}b (with control points)
and runs for the same duration.  Thus, we have realized in computation
the stabilization of the maximal mixing state over an expanded domain.

The control points need to be dynamic, following square
paths that closely track the negative defects in the original periodic
orbit.  Intuitively, the negative defects nucleate and are trapped in
the vicinity of the control points.  The positive defects, however,
are not directly controlled, but are free to move.  Note that the motion of
the control points themselves have zero topological braid entropy.
All topological entropy, and hence mixing, is generated by the response of
the positive defects.

The control points might be realized in the lab in several ways.  One
basic approach could be to use a laser to interrupt the nematic
structure at the control points, analogous to laser-melting of a
thermotropic liquid crystal~\cite{Skarabot07,Nikkhou15}.  More
generally, a variety of optical techniques have recently been
developed to control activity and guide
defects~\cite{Ross19,Zhang21b,Zarei23, Lemma23}.  Finally, one might
place at the control points a physical obstruction, such as a movable
(and ideally controllable, e.g. optically or magnetically) pillar,
large bead, or some other floating microfabricated structure, as
utilized in Refs.~\onlinecite{Ray23,Rivas20}.

The periodic behavior demonstrated here thus provides a new
possibility for taming the chaos and unpredictability of active
nematics, while also enhancing their overall mixing.

We benefitted greatly from discussions with Linda Hirst and members of
her lab at UC Merced.  This material is based upon work supported by the National
Science Foundation under Grant Nos.~DMR-1808926, DMR-2225543, and
PHY-2150531.

%


\end{document}